\documentclass[prb,preprint]{revtex4}
\usepackage{amsmath}
\usepackage{amsfonts}
\usepackage{amssymb}
\usepackage{graphicx}

\def\bea{\begin{eqnarray}}
\def\eea{\end{eqnarray}}
\def\bse{\begin{subequations}}
\def\ese{\end{subequations}}
\def\bqt{\begin{quote}}
\def\eqt{\end{quote}}
\def\nn{\nonumber}

\def\G{\Gamma}

\def\d{\delta}

\def\p{\pi}
\def\r{\rho}
\def\t{\tau}

\def\e{\epsilon}

\def\pl{\partial}
\def\ov{\over}
\def\~{\tilde}
\def\tm{\times}
\def\cd{\cdot}

\def\8{\infty}

\def\bna{\mbox{\boldmath $\nabla$}}

\def\=={\equiv}
\begin{document}

\title{Simultaneous inference of Jefimenko's and Maxwell's equations from retardation}
\author{J.-M. Chung}
\email{jmchung@khu.ac.kr}
\affiliation{Research Institute for Basic Sciences and Department of Physics,
Kyung Hee University, Seoul 130-701, Korea}


\begin{abstract}
Assuming the idea of retardation as an underlying axiom, we investigate how Jefimenko's and Maxwell's equations can be inferred. In the inference, we begin with the retarded versions of Coulomb's and Biot-Savart's field expression as an incomplete, starting ansatz. By calculating and comparing their divergences, curls, and time derivatives, we improve the ansatz. Thus improved ansatz is further improved through the same procedure and the final ansatz fields are identified with Jefimenko's fields. Our inference of Maxwell's equations is in much the same spirit as the derivation of static differential equations (divergences and curls) from the Coulomb's and Biot-Savart's fields known experimentally without knowing their governing laws of the divergence and curl equations.
\end{abstract}

\maketitle

\section{Introduction}
There have been incessant interests in the derivation of Maxwell's equations in the literature.\cite{pag,gr,za,kre,kob0,ko,kob,neu,mar,her,sch,dav,her2} The special theory of relativity, by providing the transformation properties of the electromagnetic field vectors, permits the derivation of Maxwell's equations from the field equations satisfied by a static electric field. Recent works by Davis\cite{dav} and Heras\cite{her2} provided especially nice treatments. Davis developed a generalized Helmholtz theorem for an arbitrary localized time-varying vector function and showed that the mathematical relations commonly referred to as Maxwell's equations can be derived from that theorem.
Heras formulated an existence theorem that states that, given localized scalar and vector time-dependent sources satisfying the continuity equation, there exist two retarded fields that satisfy a set of four field equations. These two formalisms are complementary.

Though Maxwell himself succeeded in establishing the system of Maxwell's equations, he did not obtain
the electric and magnetic fields in terms of the sources in general motion. Jefimenko's equations,\cite{jef1,unz} which are expressions for the electric and magnetic fields produced by time-dependent charge and current densities, have received much attention only recently\cite{gri2,ton,her4,bel,mcd} and some books\cite{hld,gri1,jac} contain a section on Jefimenko's equations in their recent editions.

The goal of this paper is to infer Jefimenko's and Maxwell's equations at the same time by adopting an appealing idea of retardation, which dictates the true nature of the generation and propagation of the fields, as a basic axiom.
The essential idea is to build up the time-dependent fields by starting from Coulomb's and  Biot-Savart's fields, and systematically improving them. Our inference of Jefimenko's equations, we believe, can be more accessible for the students of an intermediate electromagnetism course than derivations of those equations given in the literature.

The organization of the paper is as follows: In Sec.~II, we review briefly the continuity equations for sources with retarded time argument and summarize static fields and their differential equations. The starting ansatz with retarded integrals for dynamic fields is introduced Sec.~III. The detailed process of improving previously taken ansatz is described in Sec.~IV, and the identifications of Jefimenko's and Maxwell's equations are made in Sec.~V. The last section is devoted to the conclusion. In the Appendix, we list some formulas of vector analysis, especially, involving retarded source density functions.

\section{Charge Conservation at a Retarded Time}
The notions of charge density $\r$ and current density ${\bf J}$ arise from {\em one} reality,
i.e., the very existence of {\em electrically} charged particles in motion:
\bse
\bea
\r({\bf r}, t)&=&\sum q_i \d({\bf r}-{\bf r}_i(t))\,,\label{rho}\\
{\bf J}({\bf r}, t)&=&\sum q_i {d{\bf r}_i(t)\ov dt} \d({\bf r}-{\bf r}_i(t))\,.\label{j}
\eea
\ese
The conservation of charge is expressed locally by the equation of continuity at time $t'$,
\bea
\bna' \cd {\bf J}({\bf r}', t')+{\pl \r({\bf r}', t')\ov \pl t'}=0\,,\label{ce1}
\eea
whose time derivative yields another continuity equation
\bea
\bna' \cd (\pl {\bf J}({\bf r}', t')/ \pl t')+{\pl (\pl \r({\bf r}', t')/\pl t')\ov \pl t'}=0\,.\label{ce2}
\eea
The del operation ($\bna'$) in Eqs.~(\ref{ce1}) and (\ref{ce2}) pertains to the first argument (spatial argument ${\bf r}'$), not to the second argument (time argument $t'$) of the operands ${\bf J}$ and $\pl {\bf J}/ \pl t'$. We note that Eqs.~(\ref{ce1}) and (\ref{ce2}) are satisfied at any time $t'$. Thus, at a (field) retarded time $\t$ at a space point ${\bf r}'$ defined relative to (present) time $t$ at space point ${\bf r}$ by
\bea
\t\equiv t -|{\bf r}-{\bf r}'|/c\,,\label{tau}
\eea
the above continuity equations should take the following forms
\bse
\bea
\bna' \cd [{\bf J}({\bf r}', \t)]_{\t \,{\rm fixed}}+{\dot\r}({\bf r}', \t)&=&0\,,\label{cntn1}\\
\bna' \cd [{\dot {\bf J}}({\bf r}', \t)]_{\t \,{\rm fixed}}+{\ddot \r}({\bf r}', \t)&=&0\,,~~~~\label{cntn2}
\eea
\ese
where the dot over $\r({\bf r}',\t)$ and ${\bf J}({\bf r}',\t)$ denotes derivative with respect to
the second argument, $\t$. {\em Note that ${\pl f({\bf r}',\t)/ \pl \t} =\pl f({\bf r}',\t)/\pl t$ for any retarded field $f({\bf r}',\t)$, because $|{\bf r} - {\bf r'}|$ in the definition of $\t$ is a constant independent of the time $t$. This remarkable point will be exploited later frequently.}
Because of the bizarre dependence of ${\bf J}({\bf r}', \t)$ and ${\dot {\bf J}}({\bf r}', \t)$ on ${\bf r}'$, it follows that
\bse
\bea
\bna' \cd {\bf J}({\bf r}', \t)&=&-\dot\r({\bf r}', \t)+{1\ov c}{\dot {\bf J}}({\bf r}', \t)
\cd{{\bf r} - {\bf r'}\ov |{\bf r} - {\bf r'}|}\,,\label{cn1}\\
\bna' \cd {\dot{\bf J}}({\bf r}', \t)&=&-\ddot\r({\bf r}', \t)+{1\ov c}{\ddot {\bf J}}({\bf r}', \t)
\cd{{\bf r} - {\bf r'}\ov |{\bf r} - {\bf r'}|}\label{cn2}\,,
\eea
\ese
by virtue of Eqs.~(\ref{cntn1}) and (\ref{cntn2}).

In passing, the notion of time-varying density fields given by Eqs.~(\ref{rho}) and (\ref{j}) enables us to avoid a (particle) retarded time $\t_i$ endowed to each dynamic particle at the position ${\bf r}_i(\t_i)$
\bea
\t_i\equiv t -|{\bf r}-{\bf r}_i(\t_i)|/c\,.\label{taui}
\eea
As a matter of fact, due to this explicit and implicit dependence of $\t_i$ on $t$, the expressions of the Li\'{e}nard-Wiechert fields for individual charges have very complicated forms. However, by simply superposing the Li\'{e}nard-Wiechert fields due to the individual charges and introducing densities given by Eqs.~(\ref{rho}) and (\ref{j}), one might reach Jefimenko's equations.\cite{jm1}

In what follows, we will consider a system of charges in otherwise empty space; the discussion omits consideration of matter in bulk. In statics ($\dot\r=0$ and $\dot{\bf J}=0$), it is well known that the Coulomb's electrostatic field
produced by its source $\r$,
\bea
{\bf E}({\bf r})&=& {1\ov 4\p\e_0} \int_{V^\star}
{\r ({\bf r}')({\bf r} - {\bf r}')\ov |{\bf
r} - {\bf r}'|^3}d V'\,,\label{e}
\eea
and the Biot-Savart's magnetostatic field produced by its source ${\bf J}$,
\bea
{\bf B}({\bf r})&=& {1\ov 4\p\e_0 c^2} \int_{V^\star}
{{\bf J} ({\bf r}')\times({\bf r} - {\bf r}')\ov |{\bf
r} - {\bf r}'|^3}d V'\,,\label{b}
\eea
satisfy
\bse
\bea
&~&\bna\cd {\bf E}({\bf r})={1\ov \e_0}\r({\bf r})\,,\label{oe}~~~\\
&~&\bna\cd {\bf B}({\bf r})=0\,,\label{obb}\\
&~&\bna\tm {\bf E}({\bf r})=0 \label{oee}\,,\\
&~&\bna\tm {\bf B}({\bf r})={1\ov \e_0 c^2}{\bf J}({\bf r})\,.\label{ob}
\eea
\ese

In Eqs.~(\ref{e}) and (\ref{b}), the integration region $V^\star$ is all space (i.e., the entire universe).
The electric and magnetic fields given by Eqs.~(\ref{e}) and (\ref{b}) are characterized by
the superposition principle through the spacial integrations.
On one hand, we may regard the governing laws for fields [Eqs.~(\ref{oe}), (\ref{obb}), (\ref{oee}), and (\ref{ob})] as the consequences of direct applications of divergence and curl operations on empirically established physical entities [Eqs.~(\ref{e}) and (\ref{b})]. We may regard, on the other hand, the fields of Eqs.~(\ref{e}) and (\ref{b}) as the solutions to the the governing laws.

In transition to the realm of dynamical fields, authors of conventional textbooks usually introduce, in historical sequence, the empirical law of Faraday induction first and then the theoretically deduced law of Maxwell induction.
In the next section, we will attempt to modify directly the  field expressions themselves with the idea of {\em retardation} as a guiding principle, and then the governing laws, if any.

\section{A Starting Ansatz set with Retarded Fields }\label{stok}
Instantaneous action-at-a-distance extensions of Eqs.~(\ref{e}) and (\ref{b}), i.e., fields $\mathbb{E}_0({\bf r},t)$ and $\mathbb{B}_0({\bf r},t)$ obtained simply by replacing the argument ${\bf r}'$ of $\r$ and ${\bf J}$ in the integrands of Eqs.~(\ref{e}) and (\ref{b}) with $({\bf r}',t)$ do not satisfy the continuity equation, differently from the static fields of Eqs.~(\ref{e}) and (\ref{b}). We will not try other types of instantaneous action-at-a-distance extensions any more.

Let us apply this time the idea of retardation to the fields of Eqs.~(\ref{e}) and (\ref{b}) by replacing
the argument ${\bf r}'$ of source functions with $({\bf r}',t)$ to get the following starting ansatz set:
\bse
\bea
\mathbb{E}_1({\bf r},t)&=& {1\ov 4\p\e_0} \int_{V^\star}
{\r ({\bf r '},\t)({\bf r} - {\bf r'})\ov |{\bf
r} - {\bf r'}|^3}d V'\,,\label{re}\\
\mathbb{B}_1({\bf r},t)&=& {1\ov 4\p\e_0 c^2} \int_{V^\star}
{{\bf J} ({\bf r '},\t)\times({\bf r} - {\bf r'})\ov |{\bf
r} - {\bf r'}|^3}d V'\,.\label{rb}
\eea
\ese
To check whether these fields respect the law of charge conservation, we first single out $\r$ and ${\bf J}$ by taking divergence and curl on Eqs.~(\ref{re}) and (\ref{rb}), respectively. From the expression for the divergence of $\mathbb{E}_1$,
\bea
\bna\cd\mathbb{E}_1({\bf r},t)={1\ov 4\p\e_0} \int_{V^\star}
\biggl[\r ({\bf r}',\t)\bna\cd {{\bf r} - {\bf r'}\ov |{\bf
r} - {\bf r}'|^3}+{{\bf r} - {\bf r}'\ov |{\bf
r} - {\bf r}'|^3}\cd\bna \r ({\bf r}',\t)\biggr]d V'\,, \label{ree}
\eea
and Eqs.~(\ref{gr3}) and (\ref{gr6}), we obtain the charge density
\bea
\r({\bf r},t)=\e_0\bna\cd\mathbb{E}_1({\bf r},t)+{1\ov 4\p} \int_{V^\star}
{{\dot\r} ({\bf r}',\t)\ov c|{\bf
r} - {\bf r}'|^2}dV'\,.\label{de1}
\eea
We note that time derivative of this charge density, in general, does not vanish for time-varying sources, i.e.,
\bea
\dot\r({\bf r},t)\not= 0\,.\label{non}
\eea

In a similar manner, from the curl of $\mathbb{B}_1$,
\bea
\bna\tm \mathbb{B}_1({\bf r},t)&=&{1\ov 4\p\e_0 c^2}\int_{V^\star}\bigg[\biggl({{\bf r} -
{\bf r'}\ov |{\bf r} - {\bf r'}|^3} \cd \bna\biggr){\bf J} ({\bf r '}, \t)
-({\bf J} ({\bf r '}, \t)\cd \bna){{\bf r} -{\bf r'}\ov |{\bf r} - {\bf r'}|^3}\nn\\
&~&+{\bf J} ({\bf r '}, \t)\bna\cd{{\bf r} -{\bf r'}\ov |{\bf r} - {\bf r'}|^3}
-{{\bf r} -{\bf r'}\ov |{\bf r} - {\bf r'}|^3}\bna\cd{\bf J} ({\bf r '}, \t)\biggr]dV'\,,
\label{rbb}
\eea
and Eqs.~(\ref{gr3}), (\ref{gr8}), (\ref{gr12}), and (\ref{gh1}),
we obtain
\bea
{\bf J}({\bf r},t)=\e_0c^2\bna\tm \mathbb{B}_1({\bf r},t)\,,\label{de2}
\eea
where on the right-hand side a vanishing surface integral over the surface $S^\star$ surrounding the volume $V^\star$
\bea
-{1\ov 4\p}\oint_{S^\star}{{\bf J}({\bf r}',\t)(x_j-x_j')\ov |{\bf r} - {\bf r'}|^3}
\cd d{\bf A}'\label{vns1}
\eea
has been omitted.

Because the divergence of ${\bf J}$ of Eq.~(\ref{de2}) vanishes identically,
the continuity equation demand that the time derivative of $\r$ of Eq.~(\ref{de1}) also should vanish identically, which contradicts Eq.~(\ref{non}), our original assumption of time-varying source.

\section{Improvement of the starting Ansatz}\label{inf}
The origin of the dilemma arisen in the last paragraph of Sec.~III lies in that the right hand side of Eq.~(\ref{de1}) becomes $\r({\bf r},t)$, which is identical to the left-hand side of the equation.
Therefore, in order to get out of the dilemma, we should attempt to
modify the starting ansatz electric field, first, by adding $\mathbb{E}_2({\bf r},t)$ such that
\bea
\bna\cd\mathbb{E}_2({\bf r},t)={1\ov 4\p\e_0} \int_{V^\star}
{{\dot\r} ({\bf r}',\t)\ov c|{\bf r} - {\bf r}'|^2}dV'+ \G({\bf r},t)\,,\label{de111}
\eea
which absorbs the last term on the right-hand side of Eq.~(\ref{de1}) and allows an additional term $\G({\bf r},t)$. Then Eq.~(\ref{de1}) becomes
\bea
\r({\bf r},t)=\e_0\Bigl(\bna\cd[\mathbb{E}_1({\bf r},t)+\mathbb{E}_2({\bf r},t)]-\G({\bf r},t)\Bigr)\,.\label{de1111}
\eea
Reminding two formulas in vector analysis, Eqs.~(\ref{gr2}) and (\ref{va2}), one can find
\bea
\mathbb{E}_2({\bf r}, t)&\==&{1\ov 4\p\e_0} \int_{V^\star}
{\dot\r ({\bf r '}, \t)({\bf r} -{\bf r'})\ov c|{\bf r} - {\bf r'}|^2}d V'\,\label{somm}
\eea
to solve Eq.~(\ref{de111}). By virtue of Eq.~(\ref{gr7}), one can also find $\G$ to be fixed as
\bea
\G({\bf r}, t)=-{1\ov 4\p\e_0} \int_{V^\star}
{\ddot\r ({\bf r '}, \t)\ov c^2|{\bf r} - {\bf r'}|}d V'\,,\label{omm}
\eea
which can be expressed, by virtue of Eqs.~(\ref{cn1}) and (\ref{cn2}), as
\bea
\G({\bf r}, t)={1\ov 4\p\e_0}\int_{V^\star}\biggl[\bna'\cd {\dot{\bf J}}({\bf r '}, \t)
-{1\ov c}{\ddot{\bf J}}\cd{{\bf r} -{\bf r'}\ov |{\bf r} - {\bf r'}|}\biggr]{1\ov c^2|{\bf r} - {\bf r'}|}d V'\,,~~\label{ccdd}
\eea
and, further by virtue of an identity $\bna'(1/|{\bf r} - {\bf r'}|)=-\bna(1/|{\bf r} - {\bf r'}|)$ and Eq.~(\ref{gr9}) in reverse order, as
\bea
\G({\bf r}, t)=\bna\cd\biggl({1\ov 4\p\e_0}\int_{V^\star}
{{\dot{\bf J}}({\bf r '}, \t)\ov c^2|{\bf r} - {\bf r'}|}dV'\biggr)\,,\label{mx333}
\eea
where we have omitted a vanishing surface term
\bea
{1\ov 4\p\e_0c^2}\oint_{S^\star}{{\dot{\bf J}}({\bf r '}, \t)\ov |{\bf r} - {\bf r'}|}\cd d{\bf A}'\,.\label{vns3}
\eea
Eqs.~(\ref{de111}) and (\ref{mx333}) give the divergence of modified electric field $\mathbb{E}_1+\mathbb{E}_2$:
\bea
\bna\cd[\mathbb{E}_1({\bf r},t)+\mathbb{E}_2({\bf r},t)]&=&{1\ov \e_0}\r({\bf r}, t)+\bna\cd\biggl({1\ov 4\p\e_0}\int_{V^\star}
{{\dot{\bf J}}({\bf r '}, \t)\ov c^2|{\bf r} - {\bf r'}|}dV'\biggr)\,,\label{mx3}
\eea

Since we are dealing with dynamic fields, it is useful to take the time derivative,
in addition to curl and divergence, of the dynamic fields.
In these calculations, in order to relate, if possible, the electric field to the magnetic field, we will express the results in terms of ${\bf J}$ and ${\dot{\bf J}}$
by use of the charge conservation laws [Eqs.~(\ref{cn1}) and (\ref{cn2})].
By virtue of the interchangeability of two kinds of differentiations $\pl/\pl \t$ and $\pl/\pl t$ upon retarded fields, and Eqs.~(\ref{cn1}), (\ref{cn2}), (\ref{fg1}), (\ref{fg2}), (\ref{gr8}), and (\ref{gr9}) read in reverse order, we see that the time derivative of $\mathbb{E}_1+\mathbb{E}_2$,
\bea
{\pl \ov \pl t}[\mathbb{E}_1({\bf r},t)+\mathbb{E}_2({\bf r},t)]
&=&{1\ov 4\p\e_0} \int_{V^\star}\biggl[{\dot\r ({\bf r '},\t)({\bf r} - {\bf r'})\ov |{\bf
r} - {\bf r'}|^3}+{\ddot\r ({\bf r '}, \t)({\bf r} -{\bf r'})\ov c|{\bf r} - {\bf r'}|^2}\biggr]d V'
\,,\label{mm}
\eea
becomes
\bea
{\pl \ov \pl t}[\mathbb{E}_1({\bf r},t)+\mathbb{E}_2({\bf r},t)]&=&{1\ov 4\p\e_0}\int_{V^\star}\biggl[
\Bigl({\bf J} ({\bf r '}, \t)\cdot\bna\Bigr){{\bf r} -{\bf r'}\ov |{\bf r} - {\bf r'}|^3}-\Bigl(\bna\cdot{\bf J} ({\bf r '}, \t)\Bigr){{\bf r} -
{\bf r'}\ov |{\bf r} - {\bf r'}|^3}\nn\\
&~&+\Bigl({\dot{\bf J}} ({\bf r '}, \t)\cdot\bna\Bigr){{\bf r} -{\bf r'}\ov c|{\bf r} - {\bf r'}|^2}-\Bigl(\bna\cdot{\dot{\bf J}}({\bf r '}, \t)\Bigr){{\bf r} -
{\bf r'}\ov c|{\bf r} - {\bf r'}|^2}\biggr]d V'\,.~~~ \label{mmm}
\eea
On the right-hand side of Eq.~(\ref{mmm}), we have omitted a vanishing surface term
\bea
-{1\ov 4\p\e_0}\oint_{S^\star} \biggl[
{{\bf J}({\bf r}',t_r')(x_j-x_j')\ov |{\bf r} - {\bf r'}|^3}
+{{\dot{\bf J}}({\bf r}',t_r')(x_j-x_j')\ov c|{\bf r} - {\bf r'}|^2} \biggr]
\cd d{\bf A}'\,.\label{vns2}
\eea
Using Eqs.~(\ref{gr3}), (\ref{gr4}), (\ref{gr12}), (\ref{gr13}), and (\ref{va}), we eventually obtain
\bea
{\pl\ov \pl t} [\mathbb{E}_1({\bf r},t)+\mathbb{E}_2({\bf r},t)]&=&c^2\bna\tm\biggl(\mathbb{B}_1
+{1\ov 4\p\e_0 c^2}\int_{V^\star}{\dot{\bf J} ({\bf r '}, \t)\times({\bf r} -
{\bf r'})\ov c|{\bf r} - {\bf r'}|^2}d V'\biggr)\nn\\
&~&+{\pl\ov \pl t}\biggl({1\ov 4\p\e_0}\int_{V^\star}{{\dot{\bf J}} ({\bf r '}, \t)\ov c^2|{\bf r} - {\bf r'}|}d V'\biggr)
-{1\ov \e_0}{\bf J}({\bf r}, t)\,.\label{mx1}
\eea
Beginning with a correction term $\mathbb{E}_2$, we have ended up with Eq.~(\ref{mx1}), which is a natural modification of Eq.~(\ref{de2}). It can be readily shown that thus modified equations for $\r$ and ${\bf J}$ [Eqs.~(\ref{mx3}) and ~(\ref{mx1})]
do not violate the charge conservation law.

Next, let us calculate other type of spatial variation (i.e., the curl) of the modified field $\mathbb{E}_1+\mathbb{E}_2$. Relations between the spatial and time variations of the electric and magnetic fields which do not conflict with the continuity equation(s) will be the governing differential laws what we are seeking.
Using Eqs.~(\ref{gr4})--(\ref{gr7}), we see that
the expression for curl of $\mathbb{E}_1+\mathbb{E}_2$,
\bea
\bna\tm[\mathbb{E}_1({\bf r},t)+\mathbb{E}_2({\bf r},t)]&=&
{1\ov 4\p\e_0}\int_{V^\star}\bigg[\r({\bf r '}, \t)\bna\tm {{\bf r} -{\bf r'}\ov |{\bf r} - {\bf r'}|^3}
-{{\bf r} -{\bf r'}\ov |{\bf r} - {\bf r'}|^3}\tm\bna \r({\bf r '}, \t)\nn\\
&~&+\dot\r({\bf r '}, \t)\bna\tm {{\bf r} -{\bf r'}\ov c|{\bf r} - {\bf r'}|^2}
-{{\bf r} -{\bf r'}\ov c|{\bf r} - {\bf r'}|^2}\tm \bna\dot\r({\bf r '}, \t)\biggr]d V'\,,\label{aa}
\eea
results in
\bea
\bna\tm[\mathbb{E}_1({\bf r},t)+\mathbb{E}_2({\bf r},t)]=0\,.\label{mx2}
\eea

The results up to now for the divergence, curl, and time derivative of $\mathbb{E}_1+\mathbb{E}_2$ are summarized, respectively, in Eqs.~(\ref{mx3}), (\ref{mx2}), and (\ref{mx1}).

\section{Identification of Jefimenko's and Maxwell's equations}\label{inff} 
Guided by Eqs.~(\ref{mx3}) and (\ref{mx1}), we define
$\mathbb{E}_3$ and $\mathbb{B}_2$ as follows
\bse
\bea
\mathbb{E}_3({\bf r},t)&\==&-{1\ov 4\p\e_0}\int_{V^\star}{{\dot{\bf J}} ({\bf r '}, \t)\ov c^2|{\bf r} - {\bf r'}|}d V'\,,\label{e3}\\
\mathbb{B}_2({\bf r},t)&\==&{1\ov 4\p\e_0 c^2}\int_{V^\star}{\dot{\bf J} ({\bf r '}, \t)\times({\bf r} -
{\bf r'})\ov c|{\bf r} - {\bf r'}|^2}d V'\,.\label{b2}
\eea
\ese
Then, Eqs.~(\ref{mx3}) and (\ref{mx1}) can be expressed in much neater forms:
\bse
\bea
&~&\bna\cd[\mathbb{E}_1({\bf r},t)+\mathbb{E}_2({\bf r},t)+\mathbb{E}_3({\bf r},t)]={1\ov \e_0}\r({\bf r}, t)\,,\label{mxx1}\\
&~&\bna\tm[\mathbb{B}_1({\bf r},t)+\mathbb{B}_2({\bf r},t)]={1\ov \e_0 c^2}{\bf J}({\bf r}, t)+{1\ov c^2}{\pl\ov \pl t}
[\mathbb{E}_1({\bf r},t)+\mathbb{E}_2({\bf r},t)+\mathbb{E}_3({\bf r},t)]\,.~~~\label{mxx3}
\eea
\ese
To obtain an equation for the curl of the newly defined quantity $\mathbb{E}_1+\mathbb{E}_2+\mathbb{E}_3$, it should suffice to calculate
$\bna\tm\mathbb{E}_3$ because we already got Eq.~(\ref{mx2}) in hand. Also, the calculation of the divergence of the newly defined quantity $\mathbb{B}_1+\mathbb{B}_2$ will be carried out in due course.

By virtue of Eqs.~(\ref{gr1}) and (\ref{gr11}), the curl of $\mathbb{E}_3$,
\bea
\bna\tm\mathbb{E}_3({\bf r},t)&=&{1\ov 4\p\e_0}\int_{V^\star}\biggl[
{\dot{\bf J}} ({\bf r '}, \t)\tm\bna{1\ov c^2|{\bf r} - {\bf r'}|}
-{1\ov c^2|{\bf r} - {\bf r'}|}\bna\tm{\dot{\bf J}} ({\bf r '}, \t)\biggr]d V'\,,\label{ee3}
\eea
becomes
\bea
\bna\tm\mathbb{E}_3({\bf r},t)=-{1\ov 4\p\e_0c^2}\int_{V^\star}\biggl[
{{\dot{\bf J}} ({\bf r '}, \t)\tm({\bf r} - {\bf r'})\ov |{\bf r} - {\bf r'}|^3}
+{{\ddot{\bf J}} ({\bf r '}, \t)\tm({\bf r} - {\bf r'})\ov c|{\bf r} - {\bf r'}|^2}\biggr]d V'\,,
\eea
which enables us to write Eq.~(\ref{mx2}) as
\bea
\bna\tm[\mathbb{E}_1({\bf r},t)+\mathbb{E}_2({\bf r},t)+\mathbb{E}_3({\bf r},t)]
=-{\pl \ov \pl t}[\mathbb{B}_1({\bf r},t)+\mathbb{B}_2({\bf r},t)]\,.\label{mxx2}
\eea
By virtue of Eqs.~(\ref{gr4}), (\ref{gr5}), (\ref{gr10}), and (\ref{gr11}), the divergence of $\mathbb{B}_1+\mathbb{B}_2$,
\bea
\bna\cd[\mathbb{B}_1({\bf r},t)&\!\!+\!\!&\mathbb{B}_2({\bf r},t)]={1\ov 4\p\e_0c^2}\int_{V^\star}\biggl[
{{\bf r} -{\bf r'}\ov |{\bf r} - {\bf r'}|^3}\cd\bna\tm{\dot{\bf J}} ({\bf r '}, \t)
-{\bf J} ({\bf r '}, \t)\cd\bna{{\bf r} -{\bf r'}\ov |{\bf r} - {\bf r'}|^3}\nn\\
&~&~~+{{\bf r} -{\bf r'}\ov c|{\bf r} - {\bf r'}|^2}\cd \bna\tm {\dot{\bf J}} ({\bf r '}, \t)
-{\dot{\bf J}} ({\bf r '}, \t)\cd \bna\tm {{\bf r} -{\bf r'}\ov c|{\bf r} - {\bf r'}|^2}\biggr]d V'\,,\label{b5}
\eea
becomes
\bea
\bna\cd[\mathbb{B}_1({\bf r},t)+\mathbb{B}_2({\bf r},t)]=0\,.\label{mxx4}
\eea

Main results of this section are summarized by four equations: Eqs.~(\ref{mxx1}), (\ref{mxx3}), (\ref{mxx2}), and (\ref{mxx4}). It is well known that two source equations [Eqs.~(\ref{mxx1}) and (\ref{mxx3})] respect the law of charge conservation.
We remark that $\mathbb{E}_3$ and $\mathbb{B}_2$ of Eqs.~(\ref{e3}) and (\ref{b2}) have been dictated by
$\mathbb{E}_1$, $\mathbb{B}_1$, and $\mathbb{E}_2$ of Eqs.~(\ref{re}), (\ref{rb}), and (\ref{somm}); $\mathbb{E}_3$ and $\mathbb{B}_2$ are contained in Eqs.~(\ref{mx3}) and (\ref{mx1}). {\em We emphasize that $\mathbb{E}_3$ and $\mathbb{B}_2$ should not be considered to have come from the already known answer, the Jefimenko's equations.}

Identifications of $\mathbb{E}_1+\mathbb{E}_2+\mathbb{E}_3$ and $\mathbb{B}_1+\mathbb{B}_2$ with the physical electric and magnetic fields ${\bf E}$ and ${\bf B}$, i.e., Jefimenko's equations, and of Eqs.~(\ref{mxx1}), (\ref{mxx3}), (\ref{mxx2}), and (\ref{mxx4}) with Maxwell's equations complete our simultaneous inference of Jefimenko's and Maxwell's equations. Of course, the validity of these identifications should be confirmed ultimately by experiments. At the classical level, numerous experiments say that electrodynamics of Maxwell's equations (or equally Jefimenko's equations) equipped with Lorentz force law works very well.

\section{Concluding Remarks}
We have demonstrated, in detail, the process of inferring Jefimenko's and Maxwell's equations, starting from a starting  ansatz set (retarded versions of Coulomb's and Biot-Savart's fields) for time-varying fields. The method presented here is based on the knowledge of (i) the electrostatic and magnetostatic fields and the continuity equation, with which most students are familiar at the intermediate level, and (ii)
the {\em seemingly}\footnote{In reality, straightforward, but tedious! Some unfamiliar formulas, Eqs.~(\ref{gr6})--(\ref{gr13}), can be readily derived from a familiar chain rule, Eq.~(\ref{rul}).} complicated del operations on the retarded fields.

Of course, the inference of Maxwell's equations themselves here is
more complicated than the remedy by Maxwell himself of the pathology of charge non-conserving source equations.
Nonetheless, the inferring process here has its own merit. It enables us to get Jefimenko's equations with ease comparatively. Indeed, to obtain solutions to Maxwell's coupled equations except in the simplest situations is
a very complicated task. For sources in arbitrary motion, the derivation of Jefimenko's equations is known to be quite complicated; the knowledge of the generalized Helmholtz theorem (or the wave field theorem called by Jefimenko himself) or the Green function technique is prerequisite. The inferring process here, however, does not require either knowledge. In this respect, it is hoped that our inference will provide a useful tool in exposing advanced students as well as intermediate-level students to the Jefimenko's equations at early stage in the electricity and magnetism course.

\section*{Appendix: Useful formulas}
\setcounter{equation}{0}
\def\theequation{A.\arabic{equation}}
We list some vector-analysis formulas (i) not involving retarded fields
\bea
&~&\bna{1\ov |{\bf r} - {\bf r'}| }=-{{\bf r} -{\bf r'}\ov |{\bf r} - {\bf r'}|^3}\,,\label{gr1}\\
&~&\bna\cd{{\bf r} -{\bf r'}\ov |{\bf r} - {\bf r'}|^2 }={1\ov |{\bf r} - {\bf r'}|^2}\,,
\label{gr2}\\
&~&\bna\cd{{\bf r} -{\bf r'}\ov |{\bf r} - {\bf r'}|^3 }=4\p\d^3({\bf r} -
{\bf r'})\,,~~~\label{gr3}\\
&~&\bna\tm{{\bf r} -{\bf r'}\ov |{\bf r} - {\bf r'}|^2}=0\,,\label{gr4}\\
&~&\bna\tm{{\bf r} -{\bf r'}\ov |{\bf r} - {\bf r'}|^3 }=0\,,\label{gr5}
\eea
and (ii) involving retarded source density functions
\bea
&~&\bna\r({\bf r}', \t)=-{1\ov c}{\dot \r}({\bf r}', \t){{\bf r} -
{\bf r'}\ov |{\bf r} - {\bf r'}|}\,,\label{gr6}\\
&~&\bna {\dot\r}({\bf r}', \t)=-{1\ov c}{\ddot \r}({\bf r}', \t){{\bf r} -
{\bf r'}\ov |{\bf r} - {\bf r'}|}\,,\label{gr7}\\
&~&\bna \cd{\bf J}({\bf r}', \t)=-{1\ov c}{\dot {\bf J}}({\bf r}', \t)\cd{{\bf r} -
{\bf r'}\ov |{\bf r} - {\bf r'}|}\,,\label{gr8}\\
&~&\bna \cd{\dot{\bf J}}({\bf r}', \t)=-{1\ov c}{\ddot {\bf J}}({\bf r}', \t)\cd{{\bf r} -
{\bf r'}\ov |{\bf r} - {\bf r'}|}\,,\label{gr9}\\
&~&\bna \tm{\bf J}({\bf r}', \t)={1\ov c}{\dot {\bf J}}({\bf r}', \t)\tm{{\bf r} -
{\bf r'}\ov |{\bf r} - {\bf r'}|}\,,\label{gr10}\\
&~&\bna \tm{\dot{\bf J}}({\bf r}', \t)={1\ov c}{\ddot {\bf J}}({\bf r}', \t)\tm{{\bf r} -
{\bf r'}\ov |{\bf r} - {\bf r'}|}\,,\label{gr11}\\
&~&\!\!\biggl({{\bf r} -{\bf r'}\ov |{\bf r} - {\bf r'}|^3} \cd \bna\biggr){\bf J} ({\bf r '}, \t)
=-{{\dot {\bf J}}({\bf r}', \t)\ov c|{\bf r} - {\bf r'}|^2}\,,\label{gr12}\\
&~&\!\!\biggl({{\bf r} -{\bf r'}\ov |{\bf r} - {\bf r'}|^2} \cd \bna\biggr)
{\dot{\bf J}} ({\bf r '}, \t)
=-{{\ddot {\bf J}}({\bf r}', \t)\ov c|{\bf r} - {\bf r'}|}
.~~~\label{gr13}
\eea
The formulas for del operations on retarded quantities in this Appendix may be unfamiliar, but they can be derived readily from a basic differentiation rule,
\bea
{\pl f({\bf r}',\t)\ov \pl x_i}={\pl f({\bf r}',\t)\ov\pl \t}{\pl \t\ov \pl x_i}=-{1\ov c}{\pl f({\bf r}',\t)\ov\pl \t}{x_i-x_i'\ov |{\bf r}-{\bf r}|'}\,.\label{rul}
\eea

We also list other useful formulas:
\bea
&~&{\hat{\bf x}}_j\bna\cd\biggl({\bf J}({\bf r '}, \t){x_j-x_j'\ov |{\bf r} - {\bf r'}|^3}\biggr)={{\bf r} -{\bf r'}\ov |{\bf r} - {\bf r'}|^3}\bna\cd {\bf J}({\bf r '}, \t)+({\bf J} ({\bf r '}, \t)\cd \bna){{\bf r} -{\bf r'}\ov |{\bf r} - {\bf r'}|^3}\,,\label{gh1}\\
&~&{\hat{\bf x}}_j\bna'\cd\biggl({\bf J}({\bf r '}, \t){x_j-x_j'\ov |{\bf r} - {\bf r'}|^3}\biggr)
={{\bf r} -{\bf r'}\ov |{\bf r} - {\bf r'}|^3}\bna'\cd {\bf J}({\bf r '}, \t)-({\bf J} ({\bf r '}, \t)\cd \bna){{\bf r} -{\bf r'}\ov |{\bf r} - {\bf r'}|^3}\,,~~~~~~~\label{fg1}\\
&~&{\hat{\bf x}}_j\bna'\cd\biggl({\dot{\bf J}}({\bf r '}, \t){x_j-x_j'\ov |{\bf r} - {\bf r'}|^2}\biggr)
={{\bf r} -{\bf r'}\ov |{\bf r} - {\bf r'}|^2}\bna'\cd {\dot{\bf J}}({\bf r '}, \t)-({\dot{\bf J}} ({\bf r '}, \t)\cd \bna){{\bf r} -{\bf r'}\ov |{\bf r} - {\bf r'}|^2}\,,\label{fg2}\\
&~&\bna\cdot(f{\bf G})=f\bna\cdot{\bf G}+(\bna f)\cdot {\bf G}\,,\label{va2}\\
&~&\bna\tm({\bf F}\tm {\bf G})=(\bna\cdot{\bf G}){\bf F}-(\bna\cdot{\bf F}){\bf G}+({\bf G}\cdot\bna){\bf F}
-({\bf F}\cdot\bna){\bf G}\,,\label{va}
\eea

\end{document}